\documentclass[final,5p,times,twocolumn]{elsarticle}

\usepackage{amssymb}

\journal{Physics Letters A}

\begin{document}

\begin{frontmatter}

\title{Recurrence of particles in static and time varying oval
billiards}


\author{Edson D.\ Leonel$^{\rm 1}$}
\ead{edleonel@rc.unesp.br -- Phone 19-3526-9082}
\author{Carl P.\ Dettmann$^{\rm 2}$}
\address{$^{\rm 1}$Departamento de Estat\'{\i}stica, Matem\'atica
Aplicada e Computa\c c\~ao - UNESP -- Univ Estadual Paulista \\ Av. 24A,
1515 -- Bela Vista -- 13506-900 -- Rio Claro -- SP -- Brazil.\\
$^{\rm 2}$School of Mathematics, University of Bristol, Bristol BS8 1TW,
United Kingdom.}

\begin{abstract}
Dynamical properties are studied for escaping particles, injected through a hole in an oval billiard. The dynamics is considered for both static and periodically moving boundaries. For the static boundary, two different decays for the recurrence time distribution were observed after exponential decay for short times: A changeover to: (i) power law or; (ii) stretched exponential. Both slower decays are due to sticky orbits trapped near KAM islands, with the stretched exponential apparently associated with a single group of large islands. For time dependent case, survival probability leads to the conclusion that sticky orbits are less evident compared with the static case.
\end{abstract}

\begin{keyword}
Billiards; Escape of particles; Fermi acceleration

\end{keyword}

\end{frontmatter}



\section{Introduction}
\label{sec1}

Billiards are dynamical systems where an ensemble of moving particles
do not interact with each other and suffer specular reflections with the
boundary \cite{Ref1,Tab}.  Applications of billiards can be made to
different
physical systems including experiments in reflection of light from
mirrors \cite{Ref4},
superconducting \cite{Ref5} and confinement of electrons in
semiconductors by electric potentials \cite{Ref6,Ref7}, wave guides
\cite{Ref8}, microwave billiards \cite{Ref9,Ref10}, ultra-cold atoms
trapped in a laser potential \cite{Ref11,Ref12,Ref13,Ref14} and also
mesoscopic quantum dots \cite{Ref15}.
 
Billiard dynamics falls into three main classes namely: (i) regular
\cite{Tab};
(ii) intermittent \cite{Ref2} or; (iii) totally chaotic behavior
\cite{Ref3}. Two examples of case (i) are the circular billiard
\cite{Ref1}, which preserves both the energy and angular momentum, and
the elliptic billiard which preserves the energy and the product of the
angular momenta about the two foci \cite{Tab}. Case (ii) holds for
typical billiard shapes, including many models considered so far
\cite{lemon,stadium,oval,eliptico}, producing a mixed phase space
structure in the sense that elliptic islands with fractal boundaries,
generally surrounded by a chaotic sea that often is confined by
invariant spanning curves, can all be observed. The latter case
(iii) includes Sinai \cite{sinai} dispersing billiards with corners as
the diamond~\cite{BD07}, or periodically extended such as the finite
horizon Lorentz gas~\cite{D00,diego}. Billiards may also be constructed
that are fully chaotic according to the defocusing mechanism
\cite{Ref3,grigo}. However many famous examples such as the Sinai
billiard consisting of a square with a circular obstacle (or the
equivalent infinite horizon Lorentz gas~\cite{dettmann,marklof} and the
Bunimovich stadium have a regular family of periodic orbits, thus making
the dynamics intermittent as in case (ii), even though the phase space
is a single chaotic ergodic component.  This makes them analytically
tractable models for mixed systems,  similar to the more recently
introduced mushroom billiards, which have chaotic and regular regions of
phase space, but separated by smooth (rather than fractal)
boundaries~\cite{Bun01,carl_jpa}.

A time perturbation to the boundary may be considered, for example due
to
thermal vibrations in solids \cite{solids}, with amplitude and typical
frequency
related to the temperature. Depending on
the type and shape of the billiard, such a time-dependence leads to the
so called Fermi acceleration (FA) \cite{Ref19}. This phenomenon consists
in the unlimited energy growth of the bouncing particle upon collisions
with the, presumably, infinitely heavy moving boundary. Several
different kinds of perturbation can be considered in different
billiard-like models \cite{new1,new2,new3,new4,new5}. As claimed in
the Loskutov-Ryabov-Akinshin (LRA) conjecture \cite{Ref20}, if the
dynamics of the particle is chaotic while the boundary is static, thus
this is a sufficient condition to observe FA when a time perturbation to
the boundary is introduced. Recently it was shown \cite{Ref25} that even
a time-dependent elliptic billiard, which is integrable for the static
boundary, can also generate FA thanks to the appearance of a stochastic
layer replacing the separatrix curve in the phase space. Moreover, the
existence a heteroclinic orbit could extend the LRA conjecture
\cite{EDL_PRL10} and the unlimited energy growth can be observed even in
(some) integrable billiards. The occurrence of an exponential FA was
reported in a time varying rectangular billiard \cite{kedar}, which was
latter explained \cite{benno} as due to a sequence of highly correlated
motion which consists of alternating phases with free propagation motion
along the invariant spanning curves of the Fermi-Ulam model; see Ref.
\cite{EDL_JPA} for the localization of such curves for a family of
mappings whose angle is a diverging function of the action in
the limit of vanishing action, including the Fermi-Ulam model.

When a hole is introduced in the boundary therefore letting the particle
leave the billiard region, we may consider related problems of escape
(initial conditions in the billiard) or recurrence (initial conditions
at the hole); here we consider recurrence, but virtually all of the
following discussion applies also to escape, with a different exponent
in the distributions if there is a power law decay.  It is known that,
for fully chaotic dynamics, the recurrence time distribution, i.e. the
time the particle stays confined in the billiard domain, is
characterized by an exponential decay \cite{altman1}. On the other hand,
for intermittent including mixed phase space dynamics where there is
stickiness, generated from a finite time (but arbitrarily long) trapping
near periodic/elliptic regions, a power law decay is observed
\cite{altman2}. Recently, an investigation of a mushroom billiard led to
the characterization of families of marginally unstable periodic orbits
\cite{carl_jpa} responsible for trapping the particle in sticky domains,
including their effects on the escape problem. Moreover, for a time
dependent potential well \cite{diogo_pre}, the dynamics of the particle
is shown to be fully chaotic for the low energy domain and reaching
elliptic islands as far as the energy increases until finding a
limitation marked by the existence of an invariant spanning curve. For
the time dependent potential well, a hole in the energy space was
introduced letting the particle escape. Therefore for the low energy
regime an exponential decay was observed while a slower decay
characterized either as a power law or stretched exponential marks the
regime of higher energy and consequently long time.  Thus opening a
billiard by considering particles escaping through a hole is a good
means of identifying and describing various kinds of intermittency
present in the dynamics.

The oval considered here is defined by a finite Fourier series in polar
coordinates, and the mixed phase space of oval billiards was first
described by Berry in 1981 \cite{Ref2}. Since then it has remained a
popular example of a billiard with mixed phase space, including for
generalizations to time dependent boundaries \cite{FAoval} and wave
chaos in theory \cite{Sieber} and microresonator experiments
\cite{Nockel}. In this paper we revisit the oval billiard considering
both the static and time-dependent boundary.  We consider a hole for the
first time, seeking to understand and describe some properties of
particles returning to the hole.

This paper is organized as follows. In Sec. \ref{sec2}, we describe the
model with fixed boundary, detailing results for the recurrence times.
Section \ref{sec3} considers the moving boundary where the equations of
the mapping are derived. The results for the recurrence times are
discussed here also. Final remarks and conclusions are presented in Sec.
\ref{sec4}.

\section{The static oval billiard, the mapping and escaping particles}
\label{sec2}

The model we consider in this section consists of a classical
particle confined to move in a domain which the radius of the boundary
is given by the following equation in polar coordinates
\begin{equation}
R(\theta,\epsilon,p)=1+\epsilon\cos(p\theta)~,
\label{eq1}
\end{equation}
where $\epsilon$ is the amplitude of the circle's perturbation,
$\theta$ is the angular coordinate and $p>0$ is an integer. For the
parameter $\epsilon=0$ the circular billiard is obtained leading to a
foliated phase space \cite{Ref1}. Therefore chaos is not observed. In
the case of $\epsilon\ne 0$ but considering
$\epsilon<\epsilon_c=1/(p^2+1)$, the billiard is convex, and
the phase space contains both elliptic
islands, invariant spanning curves corresponding to rotating orbits
(also called whispering gallery orbits) and chaotic regions \cite{Ref17}
while for $\epsilon\ge\epsilon_c$ the billiard is no longer convex; all
the invariant tori are destroyed \cite{oval} however some elliptic
islands survive.

\begin{figure}[t]
\centerline{\includegraphics[width=1.00\linewidth]{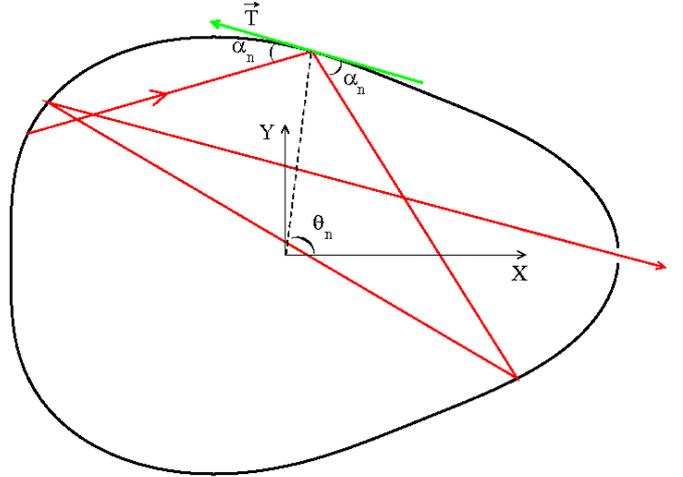}}
\caption{Illustration of the angles that describe the dynamics and an
escaping trajectory.}
\label{fig1}
\end{figure}

The dynamics of the particle is described by a two-dimensional nonlinear
area preserving map $T$ for the variables $(\theta_n,\alpha_n)$ where
$\theta_n$ is the angular position of the particle and $\alpha_n$ is the
angle that the trajectory of the particle does with respect to the
tangent vector of the boundary at the angular position $\theta_n$ (see
Fig. \ref{fig1}). The index $n$ corresponds to the $n^{th}$ collision of
the particle with the boundary. Using polar coordinates one has that $
X(\theta_n)=\left[ 1+\epsilon \cos(p \theta_n)\right]
\cos(\theta_n)$ and $Y(\theta_n)=\left[ 1+\epsilon
\cos(p\theta_n)\right] \sin(\theta_n).$ For an initial condition
$(\theta_n,\alpha_n)$, the angle between the tangent and the boundary at
the position $X(\theta_n)$ and $Y(\theta_n)$ with respect to the
horizontal is
$\phi_n=\arctan\left[Y^{\prime}(\theta_n)/X^{\prime}(\theta_n)\right]$.
Between collisions, the particle travels with a constant velocity
along a straight line until reaches the boundary. The
equation that gives the trajectory of the particle is
\begin{eqnarray}
Y(\theta_{n+1})-Y(\theta_n)=\tan(\alpha_n+\phi_n)[X(\theta_{n+1}
)-X(\theta_n)],
\label{eq5}
\end{eqnarray}
where $\phi_n$ is the slope of the tangent vector measured with respect
to the positive X-axis, $X(\theta_{n+1})$ and $Y(\theta_{n+1})$ are the
new rectangular coordinates of the collision point at $\theta_{n+1}$,
which is numerically obtained as solution of Eq. (\ref{eq5}). The angle
between the trajectory of the particle and the tangent vector to the
boundary at $\theta_{n+1}$ is
\begin{eqnarray}
\alpha_{n+1}=\phi_{n+1}-(\alpha_n+\phi_n).
\label{eq6}
\end{eqnarray}
Figure {\ref{fig1}} illustrates the corresponding angles and a escaping
particle from the billiard. The mapping that describes the dynamics of
the model is thus given by
\begin{equation}
T:\left\{\begin{array}{ll}
F(\theta_{n+1})=R(\theta_{n+1})\sin(\theta_{n+1})-Y(\theta_{n})-\\
~~~~~~\tan(\alpha_{n}+\phi_{n})[R(\theta_{n+1})\cos(\theta_{n+1}
)-X(\theta_{n})],~~\\
\alpha_{n+1}=\phi_{n+1}-(\alpha_{n}+\phi_{n})
\end{array} 
\right.
\label{eq7}
\end{equation}
where $\theta_{n+1}$ is numerically obtained as solution of
$F(\theta_{n+1})=0$ with $R(\theta_{n+1})=1+\epsilon\cos(p\theta_{n+1})$
and
$\phi_{n+1}=\arctan[Y^{\prime}(\theta_{n+1})/X^{\prime}(\theta_{n+1})]$.

\begin{figure}[t]
\centerline{\includegraphics[width=1.0\linewidth]{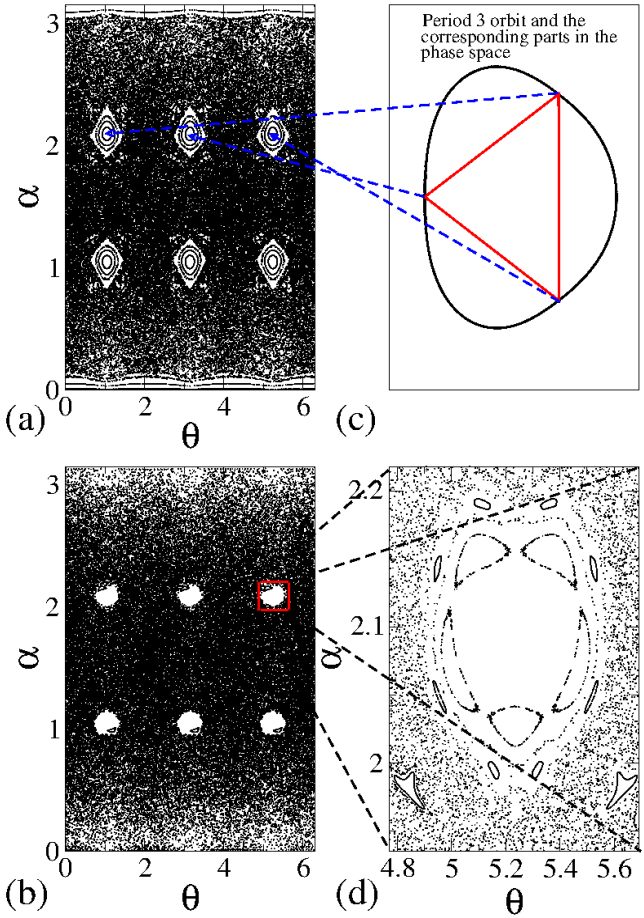}}
\caption{Phase space for the oval billiard for the control parameters:
(a) $\epsilon=0.07$ and (b) $\epsilon=0.1$. (c) shows a typical
periodic orbit and (d) a zoom-in of a specific region of (b).}
\label{fig2}
\end{figure}

A typical phase space for the static version for different control
parameters together with a visualization of a period three orbit is
shown in Fig. \ref{fig2}.
The parameters used in the figure were $p=3$ and: (a)
$\epsilon=0.07<\epsilon_c$, (b) $\epsilon=0.1=\epsilon_c$. Figure
\ref{fig2}(c) shows a period three orbit indicating corresponding
region in the phase space of (a) while (d) shows zoom-in of a region
near a elliptic island of (b).

\begin{figure}[t]
\centerline{\includegraphics[width=1.0\linewidth]{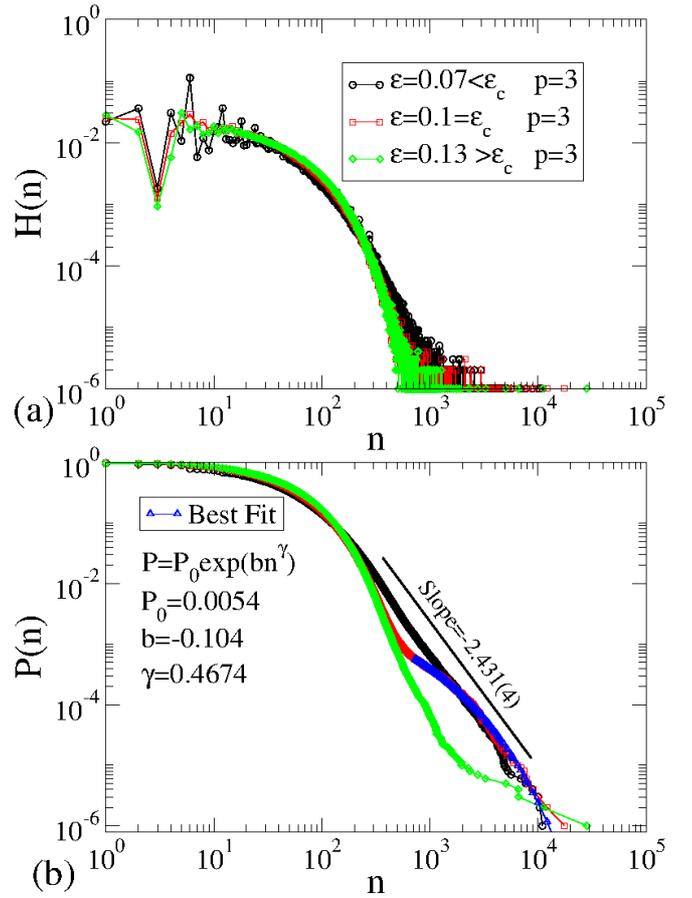}}
\caption{(a) Histogram of frequency for the escaping orbits from
the billiard. (b) Corresponding survival probability, obtained by
integration of the histogram shown in (a). The parameters used were
$p=3$ and $\epsilon=0.07<\epsilon_c$, $\epsilon=0.1=\epsilon_c$ and
$\epsilon=0.13>\epsilon_c$.}
\label{fig3}
\end{figure}

\begin{figure}[t]
\centerline{\includegraphics[width=1.0\linewidth]{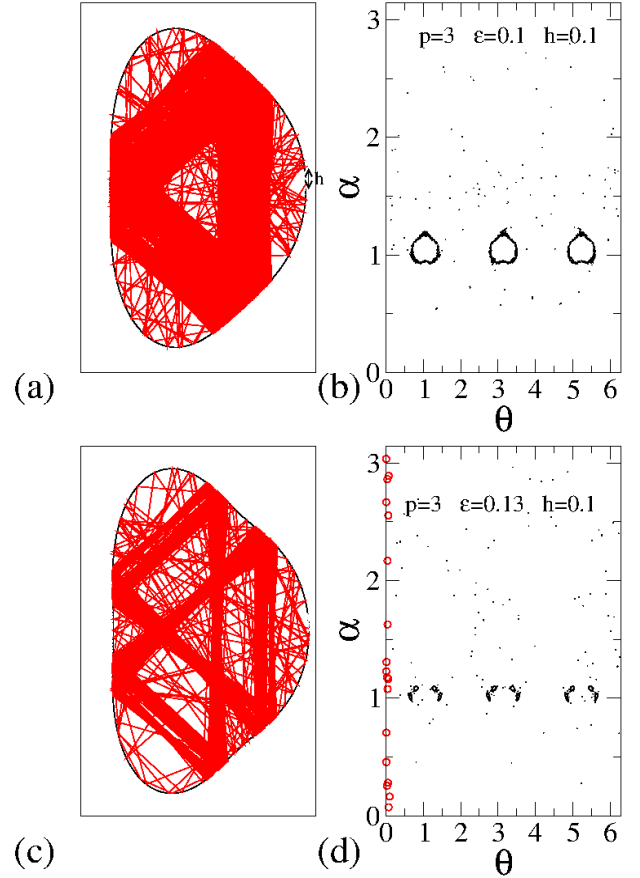}}
\caption{Plot of a typical long-lived orbit in the billiard (a,c) and
its corresponding representation in the phase space (b,d) for the
parameters $p=3$ $h=0.1$ and: (a,b) $\epsilon=0.1=\epsilon_c$; (c,d)
$\epsilon=0.13>\epsilon_c$.}
\label{fig_tra}
\end{figure}

Let us now consider that the boundary has a hole through which the
particles are injected and can escape, as shown in Fig. \ref{fig1}. We
assume the hole is localized in $\theta\in(0,h)$ where $h$ is a
parameter. We simulated different values of $h\le\pi/10$
however in this paper we fix it at $h=0.1$. The procedure used to
consider the escape of the particles assumes the evolution of an
ensemble of initial conditions. Indeed we consider $10^6$  different
initial conditions in a window where $10^3$ $\theta_0$ are uniformly
distributed along $\theta_0\in(0,h)$ while a window of $10^3$ different
$\alpha_0$ also uniformly distributed along $\alpha_0\in (0,\pi)$. Each
one of them was let to evolve a maximum of $10^6$ collisions with the
boundary, if it did not escape before. When the particle reaches the
region of the hole for the first time, the number of collisions with the
boundary spent up to that point is registered and the particle is
assumed to escape. A new initial condition is then started and the
procedure is repeated until all the  ensemble is exhausted. The
histogram of frequency of escaping particles, represented as $H(n)$ is
shown in Fig. \ref{fig3}(a) for three different parameters, as labeled
in the figure.
For a fixed $p=3$, the parameter $\epsilon=0.07<\epsilon_c$ causes the
phase space to have both elliptic islands as well as invariant spanning
curves \cite{oval} corresponding to the so called whispering gallery
orbits. The presence of the elliptic islands leads the dynamics of
some initial conditions to experience a sticky behavior that can be
long. The invariant spanning curves are destroyed for the cases of
$\epsilon=0.1$ and $\epsilon=0.13$. Figure \ref{fig3}(a) shows the
behavior of the histogram of escaping orbits. The horizontal axis
denotes the number of collisions the particle suffered with the boundary
before escaping while the vertical one corresponds to the fraction of
orbits which escaped at the $n^{\rm th}$ collision. Given that $p=3$ one
can see that the histogram shows a lower value for three bounces with
boundary as compared with $n=2$ and $n=4$. This reduction is related to
the stability of period three orbits in the phase space, therefore
trapping the particle close to this region. For large $n$ we see a long
tail which corresponds to sticky orbits. The integration of such
histogram gives the so called cumulative recurrence time distribution,
which is defined as
\begin{equation}
P={1\over N}{\sum_{j=1}^N{N_{\rm rec}(n)}}~,
\label{def}
\end{equation}
where the summation is taken along the ensemble of $N=10^6$ different
initial conditions. $N_{\rm rec}(n)$ denotes the number of initial
conditions that do not escape through the hole (ie recur) until a
collision $n$. When Eq. (\ref{def}) is evaluated in a fully chaotic
dynamics its behavior is an exponential \cite{altman1} while for a mixed
phase space where intermittent orbits exist along the phase space a
power law is observed \cite{altman2}. We have shown recently \cite{carl}
that the existence of elliptic islands may also lead to a stretched
exponential decay. Figure \ref{fig3}(b) shows the behavior of three
curves of $P(n)~vs~n$ for the same set of control parameters used in
Fig. \ref{fig3}(a). For $\epsilon=0.07<\epsilon_c$ the decay is
exponentially fast at the beginning until about $200$ collisions of the
particle with the boundary when the curve changes to a slower decaying
regime marked by a power lay with exponent $-2.431(4)$. For
$\epsilon=0.1=\epsilon_c$, the invariant spanning curves creating the
whispering gallery orbits are destroyed. The decay of $P$ at the
beginning is the same for $\epsilon=0.07$ when a hump appeared around
$n\cong 500$ lasting until $n\cong 1500$. Indeed the hump is described
by a stretched exponential of the type
\begin{equation}
P=P_0\exp(b~n^\gamma)
\label{P}
\end{equation}
with the coefficients $P_0=0.0054$, $b=-0.104$ and $\gamma=0.4674\cong
0.5$. From the $10^6$ different initial conditions, the region
corresponding to the hump is due to $947$ initial conditions. The major
part of the trapping happens near a period three orbit as shown in Fig.
\ref{fig_tra}(a) with the corresponding sticky orbit plotted in Fig.
\ref{fig_tra}(b).
Finally, for $\epsilon=0.13>\epsilon_c$ the elliptic regions in the
phase space are reduced and the recurrence distribution decays rapidly
at first. After $n>1360$, some of the initial conditions are trapped for
long time in a sticky region. The few orbits trapped for a long time
were mostly observed near a period twelve orbit, as shown in Fig.
\ref{fig_tra}(c) with its corresponding plot in the phase space shown in
Fig. \ref{fig_tra}(d).

\section{Time dependent oval billiard and escaping particles results}
\label{sec3}

This section is devoted to discussing the recurrence time distribution
of the time-dependent oval billiard. We first construct the mapping that
gives the precise description of the dynamics. The radius of the
boundary in polar coordinates, to include the time-dependence, is now
written as
\begin{equation}
R_b(\theta,t)=1+\epsilon[1+a\cos(t)]\cos(p\theta)~,
\label{rf_t}
\end{equation}
where $a$ corresponds to the amplitude of oscillation of the boundary.
The introduction of the time perturbation to the boundary produces
two new additional variables that have to be considered, namely: (i) the
velocity of the particle, $V$, and the time $t$. The map describing
the dynamics has now four dynamical variables, i.e.
$T(\theta_n,\alpha_n,V_n,t_n)=(\theta_{n+1},\alpha_{n+1},V_{n+1},t_{n+1}
)$. Supposing the initial conditions $(\theta_n,\alpha_n,V_n,t_n)$ are
given, a similar procedure as made in the previous section can be used
to describe the position and trajectory of the particle. Then the
instant of the collision is obtained by the numerical solution of the
following equation
\begin{equation}
R_p(\theta,t)=R_b(\theta,t)~,
\label{r_p=r_f}
\end{equation}
where $R_b(\theta,t)=1+\epsilon[1+a\cos(t_n+t)]\cos(p\theta_p)$ and
$R_p(t)=\sqrt{X^2_p(t)+Y^2_p(t)}$ with the corresponding angle
$\theta_p=\arctan[Y_p(t)/X_p(t)]$,
$X_p(t)=X(\theta_n,t_n)+|\vec{V}_n|\cos(\phi_n+\alpha_n)(t-t_n)~,
\label{x_p}$ and $
Y_p(t)=Y(\theta_n,t_n)+|\vec{V}_n|\sin(\phi_n+\alpha_n)(t-t_n)~,
\label{y_p}$ with $t\ge t_n$.

The angular coordinate at the new collision, $\theta_{n+1}$, is
numerically obtained from  Eq. (\ref{r_p=r_f}) via a numerical
procedure similar to the molecular dynamics method leading to an
accuracy of $10^{-12}$ in the time of the collision. Given
$\theta_{n+1}$, the time at the collision is written as
\begin{equation}
t_{n+1}=t_n+{{[(\Delta X)^2+ (\Delta Y)^2]^{1/2}}\over{v_n}}~.
\label{tempo}
\end{equation}
where $\Delta X=X_p(\theta_{n+1})-X_p(\theta_n)$ and $\Delta
Y=Y_p(\theta_{n+1})-Y_p(\theta_n)$. The velocity of the moving boundary
is given by
\begin{eqnarray}
\vec{v}_b(t_{n+1})&=&-\epsilon a\sin(t_{n+1})\cos(p\theta_{n+1})
\nonumber\\
&\times& [\cos(\theta_{n+1})\vec{i}+\sin(\theta_{n+1})\vec{j}]~.
\label{v_f}
\end{eqnarray}

\begin{figure}[t]
\centerline{\includegraphics[width=1.0\linewidth]{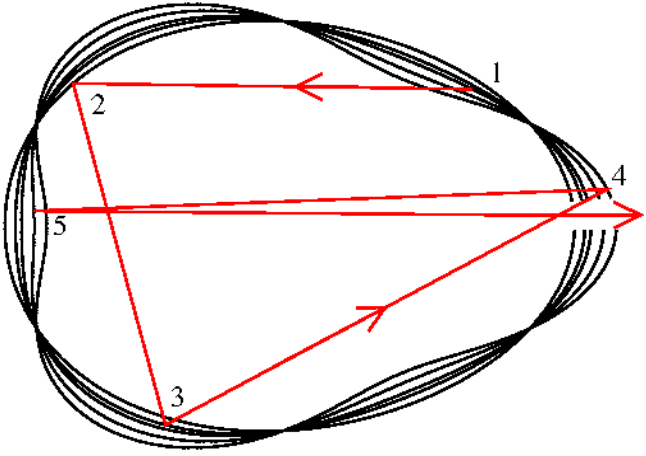}}
\caption{Plot of five snapshots of a trajectory for an escaping particle
in the time varying billiard. The control parameters used, for visual
purposes, were $p=3$, $\epsilon=0.1$ and $a=0.8$.}
\label{fig4}
\end{figure}

\begin{figure}[t]
\centerline{\includegraphics[width=1.0\linewidth]{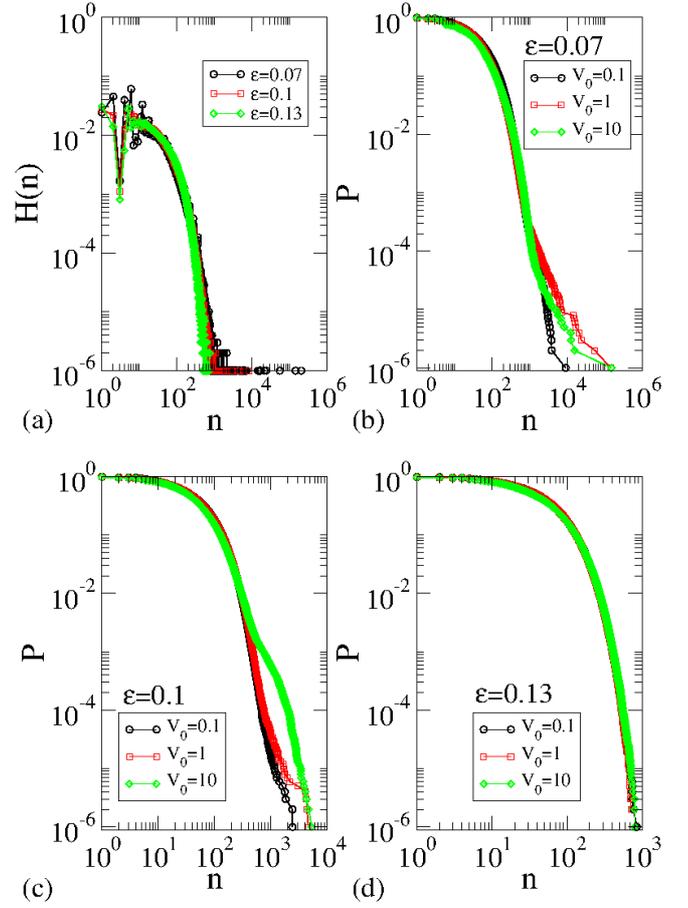}}
\caption{(a) Histogram of frequency for the escaping orbits from
the billiard. (b) Corresponding survival probability, obtained by
integration of the histogram shown in (a). The parameters used were
$p=3$ and $\epsilon=0.07<\epsilon_c$, $\epsilon=0.1=\epsilon_c$ and
$\epsilon=0.13>\epsilon_c$ and $a=0.1$.}
\label{fig5}
\end{figure}

The reflection laws used are
\begin{eqnarray}
\vec{V}^{\prime}_{n+1}\cdot \vec{T}_{n+1}=\vec{V}^{\prime}_n\cdot
\vec{T}_{n+1}~,\label{c1}\\
\vec{V}^{\prime}_{n+1}\cdot \vec{N}_{n+1}=-\vec{V}^{\prime}_n\cdot
\vec{N}_{n+1}~,
\label{c2}
\end{eqnarray}
where the upper prime denotes the variables are represented in the
moving referential frame. From Eq. (\ref{c1}) one concludes that the
tangent component of the velocity does not indeed suffers any
modification after the impact. Returning to the inertial frame of
reference, we obtain that
\begin{eqnarray}
\vec{v}_{n+1}&\cdot&
\vec{T}_{n+1}=v_n[\cos(\alpha_n+\phi_n)\cos(\phi_{n+1})]+\nonumber\\
&+&v_n[\sin(\alpha_n+\phi_n)\sin(\phi_{n+1})]~.
\label{vt}
\end{eqnarray}
Considering Eq. (\ref{c2}), in the rest referential frame, the normal
component of the velocity of the particle is
\begin{eqnarray}
\vec{v}_{n+1}&\cdot&
\vec{N}_{n+1}=-v_n[-\cos(\alpha_n+\phi_n)\sin(\phi_{n+1})]+\nonumber\\
&+&v_n[\sin(\alpha_n+\phi_n)\cos(\phi_{n+1})]-\nonumber \\
&-&2\epsilon
a\sin(t_{n+1})\cos(p\theta_{n+1})[-\cos(\theta_{n+1})\sin(\phi_{n+1})]
-\nonumber \\
&-&2\epsilon a\sin(t_{n+1})\sin(\theta_{n+1})\cos(\phi_{n+1})]~.
\label{vn}\nonumber
\end{eqnarray}
The velocity of the particle immediately after the impact is
\begin{equation}
v_{n+1}=\sqrt{(\vec{v}_{n+1}\cdot
\vec{T}_{n+1})^2+(\vec{v}_{n+1}\cdot\vec{N}_{n+1})^2}~,
\label{v1}
\end{equation}
and finally, the angle that the particle leaves the boundary, measured
with respect to a tangent to the point $\theta_{n+1}$ is written as
\begin{equation}
\alpha_{n+1}=\arctan\left[{{\vec{v}_{n+1}\cdot\vec{N}_{n+1}}
\over{\vec{v}_{n+1}\cdot\vec{T}_{n+1}}}
\right]~.
\label{alpha1}
\end{equation}

For $a\ne 0$ the particle can gain or lose energy upon collisions with
the boundary and given that the phase space has chaotic components,
unlimited energy growth is observed \cite{Ref21} therefore
confirming the LRA conjecture. Figure \ref{fig4} shows 5 snapshots of an
orbit as well as the corresponding position of the wall at the instant
of the collisions. The parameters used, only for visual purposes were
$p=3$, $\epsilon=0.1$, $a=0.8$ with $h=0.1$.

The histogram of frequency for the escaping orbits is shown in Fig.
\ref{fig5}(a) for the parameters $p=3$, $a=0.1$ and the same three
different $\epsilon$, as used in the static case. We can see from the
histogram (Fig. \ref{fig5}(a)) that the escaping particles at 3
collisions are still less observed than the ones for 2 and 4 collisions.
The survival probability was considered for different values of
$\epsilon$ and for three different values of initial velocity. For
$\epsilon=0.07$, we can see in Fig. \ref{fig5}(b) that the survival
probability decays exponentially fast and few orbits keep trapped at
large $n$ therefore leading the curve to slower the decay at the end.
The slower initial velocity $V_0=0.1$ seems to affect less the dynamics
while a short tail of slower decay is observed for $V_0=1$ and
$V_0=10$. For $\epsilon=0.1$ (see Fig. \ref{fig5}(c)), the survival
probability for $V_0=10$ is marked by a hump starting at $n\approx 400$
while few orbits are trapped in sticky domain for $V_0=1$ and
$V_0=0.1$. On the other hand, when $\epsilon=0.13$ (see Fig.
\ref{fig5}(d)), no significant changes from fast exponential decay was
observed as dependent on the initial velocity. 

Let us discuss the results obtained for the survival probability
considering the two cases of (i) static and (ii) time-varying boundary.
For the static boundary, the velocity of the particle is constant and
for a mixed phase space where invariant spanning curves and KAM islands
coexist, the sticky behavior observed is larger as compared to the case
where invariant spanning curves are destroyed ($\epsilon\ge\epsilon_c$
(see Ref. \cite{oval})) and chaotic sea is limited by KAM islands only.
Therefore the decay of the survival probability starts exponentially
at short collisions and suddenly it is marked by a changeover turning
into a power law for large number of collisions. On the other hand, when
the invariant spanning curves are destroyed but the period three region
in the phase space still influences the dynamics, several instances of
trapping were
observed leading the dynamics to spend long time near period three
orbits. The decay of the survival probability starts exponentially fast
at short collisions and suddenly it changes to a slower decay being
characterized by a stretched exponential. The decay is slower than
exponential but is still faster than a power law. Indeed, as the control
parameters are varied and invariant spanning curves are destroyed, one
can observe a continuum spectrum of decay ranging from exponential to a
power law. This variation is still an open problem and extensive
theoretical
and numerical investigations should be made to describe it properly. As
the
parameter $\epsilon$ rises, the elliptic region in the phase space
decreases and the exponentially fast decay of the survival probability
is most evident. However trapping is still observed for large
times. In our case we observed a few long orbits trapped near a
region of period twelve. Such orbits indeed slow the decay of the
survival probability at the very long time but were observed only in a
few
trajectories.

For case (ii) where a time varying boundary is considered, the
velocity of the particle is no longer constant. The LRA
conjecture claims \cite{Ref20}, the chaotic dynamics of the particle for
the static boundary leads to the unlimited energy growth when a time
perturbation is considered; numerical studies of this model are
consistent with this prediction \cite{FAoval,Ref21}. A consequence is
that a
particle with high energy collides many more times with the boundary in
a given interval of time while compared with a lower energy particle at
the same interval of time.  Over a small number of collisions the
billiard
it sees is effectively static, and it is likely to escape well before
Fermi
acceleration is evident. This is observed particularly for the case of
$\epsilon=0.1$ and $V_0=10$ where a hump in the survival probability
is evident.

\section{Concluding remarks}
\label{sec4}

We have studied some dynamical properties of an oval-like billiard
with a hole in the boundary, considering both static as well as time
dependent boundaries. For the static case, the recurrence time
distribution of the hole has a fast decay for short collisions changing
the decay either to a power law or stretched exponential, depending on
the control parameter. The power law observed for $\epsilon=0.07$ in
the static case has slope $-2.431(4)$ while the stretched exponential
for $\epsilon=0.1$ is given by $P=P_0\exp(b~n^\gamma)$ with coefficients
$P_0=0.0054$, $b=-0.104$ and $\gamma=0.47$. The sticky orbits present in
the dynamics are responsible for slowing the decay of the recurrence
time distribution. For the time dependent case, the survival
probability was considered for different values of $\epsilon$ and for
three different values of initial velocity. For $\epsilon=0.07$, the
lower initial velocity seems to affect less the trapping orbits while a
short tail is observed for larger initial velocities indicating a sticky
regime. For $\epsilon=0.1$, the initial $V_0=10$ lead the survival
probability to exhibit a hump starting at $n\approx 400$ therefore
indicating a sticky regime. Indeed at that large energy, the particle
suffers many more collisions with the boundary at the same
interval of time as compared to a low energy particle, hence seeing
less the influence of the
moving boundary compared with a lower energy particle, therefore
seem to be more susceptible to sticky behavior. For $\epsilon=0.13$ the
initial velocities considered do not seem to change significantly the
fast exponential decay as observed for $\epsilon=0.1$ and
$\epsilon=0.07$.

The observation of stretched exponential decays, as in Ref.~\cite{carl},
invites further investigation, as previous studies have concentrated on
algebraic decay models.  We would like to remark that as in
Ref.~\cite{carl},
a stretched exponential decay is observed where there is a single
prominent elliptic periodic orbit responsible for the stickiness, and
also
that the exponent $\gamma$ is very close to $1/2$.  In general, opening
a dynamical system with a hole is a very effective method of elucidating
the structure of intermittent dynamics.

\section*{Acknowledgments}
EDL acknowledges support from CNPq, FAPESP and FUNDUNESP, Brazilian
agencies. CPD is grateful to PROPe-UNESP for his visit to
DEMAC-UNESP/Rio Claro. This research was supported by resources supplied
by the Center for Scientific Computing (NCC/GridUNESP) of the S\~ao
Paulo State University (UNESP).

\end{document}